\documentclass[twocolumn,a4paper,showpacs,pra,eqsecnum]{revtex4}

\usepackage{epsfig,amssymb}

\begin{document}
\newcommand{\eqr}[1]{Eq.~(\ref{#1})}
\newcommand{\id}{\openone}
\newcommand{\beq}{\begin{equation}}
\newcommand{\eeq}{\end{equation}}
\newcommand{\beqa}{\begin{eqnarray}}
\newcommand{\eeqa}{\end{eqnarray}}
\newcommand{\beqan}{\begin{eqnarray*}}
\newcommand{\eeqan}{\end{eqnarray*}}
\newcommand{\nn}{\nonumber}
\newcommand{\nl}[1]{\nn \\ && {#1}\,}
\newcommand{\erf}[1]{(\ref{#1})}
\newcommand{\erfs}[2]{Eqs.~(\ref{#1})--(\ref{#2})}
\newcommand{\crf}[1]{Ref.~\cite{#1}}
\newcommand{\dg}{^\dagger}
\newcommand{\rt}[1]{\sqrt{#1}\,}
\newcommand{\smallfrac}[2]{\mbox{$\frac{#1}{#2}$}}
\newcommand{\half}{\smallfrac{1}{2}}
\newcommand{\bra}[1]{\langle{#1}|}
\newcommand{\ket}[1]{|{#1}\rangle}
\newcommand{\ip}[2]{\langle{#1}|{#2}\rangle}
\newcommand{\braket}[2]{\langle{#1}|{#2}\rangle}
\newcommand{\sch}{Schr\"odinger}
\newcommand{\hei}{Heisenberg}
\newcommand{\ito}{It\^o}
\newcommand{\str}{Stratonovich}
\newcommand{\bl}{{\bigl(}}
\newcommand{\br}{{\bigr)}}
\newcommand{\dbd}[1]{{\partial}/{\partial {#1}}}
\newcommand{\sq}[1]{\left[ {#1} \right]}
\newcommand{\cu}[1]{\left\{ {#1} \right\}}
\newcommand{\ro}[1]{\left( {#1} \right)}
\newcommand{\an}[1]{\left\langle{#1}\right\rangle}
\newcommand{\tr}[1]{{\rm Tr}\sq{ {#1} }}
\newcommand{\du}{\partial}
\newcommand{\strike}[1]{{#1}\vspace{-3.5ex} \\ {\bf \underline{\phantom{#1}}}}
\newcommand{\tp}{^{\top}}
\newcommand{\tbt}[4]{\left( \begin{array}{cc} {#1}& {#2} \\ {#3}&{#4} \end{array}\right)}
\newcommand{\proj}[1]{\ket{#1}\bra{#1}}

\newcommand{\Tr}{{\rm Tr}}
\newcommand{\A}{_{\rm A}}
\newcommand{\B}{_{\rm B}}
\newcommand{\AB}{_{\rm AB}}
\newcommand{\p}{_{\rm P}}
\newcommand{\m}{_{\rm M}}
\newcommand{\n}{_{\rm N}}
\newcommand{\N}{^{(N)}}
\newcommand{\M}{^{(M)}}
\newcommand{\ary}[2]{\begin{array}{#1}  #2  \end{array}}
\newcommand{\U}{{\rm U}}

\setlength\arraycolsep{1pt}

\title{Optimal reference states for maximum accessible entanglement under
the local particle number superselection rule}
\author{G.A. White}
\author{J.A. Vaccaro}
\author{H.M. Wiseman}
\affiliation{Centre for Quantum Computer Technology, Centre for
  Quantum Dynamics, School of Science, Griffith University, Brisbane,
  4111 Australia}

  \date{\today}

\begin{abstract}
Global conservation laws imply superselection rules (SSR) which
restrict the operations that are possible on any given state.
Imposing the additional constraint of local operations and classical
communication (LOCC) forbids the transfer of quantum systems between
spatially separated sites. In the case of particle conservation this
imposes a SSR for {\em local} particle number. That is, the
coherences between subspaces of fixed particle number at each site
are not accessible and any state is therefore equivalent to its
projection onto these subspaces. The accessible entanglement under
the SSR is less than (or equal to) that available in the absence of
the SSR. An ancilla can be used as a reference system to increase
the amount of accessible entanglement. We examine the relationship
between local particle number uncertainty and the accessible
entanglement and consider the optimal reference states for
recovering entanglement from certain systems. In particular we
derive the optimal ancilla state for extracting entanglement for a
single shared particle and make steps towards the optimum for
general systems. We also show that a reference for phase angle is
fundamentally different to a reference for the SSR associated with
particle conservation.
\end{abstract}

\pacs{03.67.-a, 03.67.Mn, 11.30.-j, 03.65.Ta}
%

\maketitle

\section{Introduction}\label{Intro}
Some five decades ago Wick, Wightman and Wigner showed that conservation laws induce so called ``superselection
rules'' (SSR) that forbid the observation of coherences between different eigenstates of the conserved quantity by any
physical measurement \cite{WWW}.  A decade later Aharanov and Susskind demonstrated that this restriction can be
alleviated with an additional system that acts as a reference for the coherences \cite{Suss}. Such reference ancillae are of
particular interest when the dimension of the Hilbert space it occupies is not arbitrarily large compared to that of the
system. In such cases the quantum nature of the reference system is important.  We refer to such a reference system as a
{\it finite reference} because of the finite dimension of the reference system's Hilbert space \footnote{ Our definition of
{\it finite} reference systems differs from the {\it bounded} reference systems of Bartlett {\em et al.} \cite{review}.  In
the latter case the bound is on some physical parameter such as the number of spin 1/2 particles or the mean photon
number. In particular, bounded reference systems can have either finite or infinite dimensional Hilbert spaces. In contrast
finite reference systems are restricted by having a finite dimensional Hilbert space.}. This is the regime that we are
mainly concerned with here.

There has been some laxity in the use of terms to distinguish the classical and quantum regimes in the literature. To
avoid any possible confusion we shall reserve the term {\it reference frame} for a set of independent variables which, in
principle, can be defined without an explicit relation to any physical object. The inertial reference frames in special
relativity are an example.  In contrast, we use the terms {\it reference system} and {\it reference ancilla} to describe a
physical system whose observables are used as a reference for other physical systems. This distinction between frames
and systems is crucial in theories such as quantum gravity where external reference frames are artificial and a description
of a physical system can only be made relative to other physical systems \cite{poulin1,poulin2}.

The lack of the ability to observe coherences under a SSR implies that there are operations whose effect on the system
are not physically detectable.  Conversely, the existence of such operations implies that a SSR operates. This converse
situation is true even in the absence of a conservation law. For example, it can be shown that the set of non-detectable
unitary operations form a symmetry group, $G$ \cite{VAWJ}. In other words, $G$ is the symmetry group which
describes the invariance of the system under the SSR. Moreover, this invariance implies that the set of physically allowed
operations $\{\cal O(\cdot)\}$ is constrained to commute with the symmetry group $G$ \cite{BartlettWiseman}, that is,
\beq
             {\cal O} [\hat{T} (g) \hat\rho \hat{T}^\dagger (g) ] = \hat{T} (g){\cal O}  [\hat\rho ] \hat{T} ^\dagger(g)\quad \forall g\in G\  ,
            \label{O commutes with G}
\eeq
for every state $\hat\rho$ of the system. Here $\hat{T}(g)$ is the unitary representation of the transformation $g \in G$.
In general, ${\cal O}$ represents an operation on an open system and the openness implies that any global conservation
law may not hold for the system itself. This shows that SSRs can exist independently of conservation laws at least for
open systems.

In accord with previous work, we label the SSR associated with the symmetry group $G$ as the $G$-SSR. Alternatively,
if $G$ is characterized by a generator, we sometimes refer to the SSR by the name of the generator. For example,  a
unitary representation of U(1) for a system of identical particles is given by $\{e^{i\hat N\theta}:0\le\theta < 2\pi\}$
where the group generator $\hat N$ is the particle number operator, and so here the U(1)-SSR is equivalent to the
particle number SSR.

The situation is further enriched if, in addition to a SSR, we also impose the  constraint of allowing only local operations
and classical communication (LOCC) for a bipartite system.  The physically allowed operations $\cal O$ must satisfy
both \eqr{O commutes with G} and LOCC. Consider the bipartite case with subsystems at two spatially separated sites A
and B. Any local operation ${\cal O}={\cal O}_A\otimes{\id}_B$ that satisfies \eqr{O commutes with G} has the
property that
\beq
         ({\cal O}_A\otimes{\id}_B) [\hat{T} (g) \hat\rho \hat{T}^\dagger  (g)]=
             \hat{T} (g) ({\cal O}_A\otimes{\id}_B)[\hat\rho ] \hat{T} ^\dagger(g)
\eeq for all $g\in G$, where $\cal O_\mu$ is a local operator acting
at site $\mu$ and $\id$ is the identity operator. If the group has
local representations $\{\hat{T}_\mu (g) :g \in G \}$ for each site
$\mu$ where $\hat{T}(g) = \hat{T}_A(g)\otimes \hat{T} _B(g)$ then
\beq
     {\cal O}_\mu [\hat{T}_\mu (g) \hat\rho \hat{T}^\dagger _\mu (g) ]=
         \hat{T}_\mu  (g) {\cal O}_\mu [\hat\rho ] \hat{T}_\mu ^\dagger (g)\quad \forall g\in G\ .
\eeq
For the case of a Lie group, this requires its local generators to be additive. Thus the combination of LOCC and a SSR
can induce a local SSR acting at each site. Such local SSRs are not necessarily fundamental in the sense of arising from a
conservation law like that of charge, but are {\em imposed} by our interest in examining the effect of the LOCC
constraint. Essentially, a local SSR arises when two sites lack a shared reference to break the symmetry represented by
$G$ at each site \cite{BartlettWiseman,HMW,VacAnsWis,VAWJ}.  However, the absence of the shared reference here
is due to the LOCC constraint rather than any fundamental constraint. For example, the local U(1)-SSR due to the
absence of a shared optical phase reference is imposed by the prohibition on transporting quantum fields between two
sites under the LOCC constraint rather than the conservation of photon number.

Interest in superselection rules and their associated reference systems has revived in recent years particularly in relation
to quantum information \cite{review}.   Bartlett {\it et al.} have shown that in the absence of a reference for spatial
orientation the communication of classical and quantum information using spin particles is still possible provided the
spins are entangled and the information is encoded in SSR-invariant subsystems \cite{BRS-lack-ref}.  They have also
shown how a private shared reference for spatial orientation can be used for secret communications
\cite{BRS-private-ref}. The effects of finite references has been studied in a variety of ways.

For example, Bartlett {\it et al.} \cite{BRS-lack-ref}, Bagan {\it et al.} \cite{Bagan} and Lindner {\it et al.}
\cite{Lindner} have examined protocols for estimating the relative angle between two directions which are defined by
finite dimensional spin systems. Further, the effect on quantum operations due to finite references was studied initially
by van Enk and Kimble \cite{vanEnkKimble} and shortly later by Gea-Banacloche \cite{Gea-Banacloche} and, although
specific details of this initial work attracted some criticism from Itano \cite{Itano} and Nha  and Carmichael
\cite{Carmichael}, it has been extended and generalized to quantum measurements, uncertainty relations and
simultaneous measurements mainly by Ozawa \cite{Ozawa1,Ozawa2,Ozawa3,Ozawa4}.

Entanglement is also affected by the presence of a SSR.  A SSR has an effective decohering effect that constrains the
amount of entanglement in a system that is accessible.  A number of different terms have been used to label the
entanglement which is constrained in this way. Bartlett and Wiseman \cite{BartlettWiseman}, as well as others (see e.g.
\cite{VAWJ,VacAnsWis,steve,Dowling}), refer to it as the accessible entanglement \footnote{We note that Jones {\em
et al.} \cite{steve} use the term {\it extractable entanglement} in place of accessible entanglement.}. For the special case
of indistinguishable particles in the absence of a shared phase reference, we previously referred to the SSR-constrained
entanglement as particle entanglement in Refs.~\cite{HMW,VacAnsWis} to distinguish it from the entanglement of the
spatial modes occupied by the particles. Our operational definition of particle entanglement makes use of a set of local
quantum ancillary systems (or registers) which are not subject to the SSR. The accessible entanglement in a system of
shared identical particles is given by the maximum amount of entanglement that one can transfer from the system to the
local ancillas by $G$-invariant operations that obey LOCC. It is important to note that the transfer operation involves
only a {\it single} copy of the system state. However, once the entanglement resides in the ancillae it can be manipulated
in the usual way free of the SSR. Further details of this non-asymptotic interpretation of particle entanglement can be
found in the Introduction of Ref.~\cite{VAWJ}.

In this paper we focus on the problem of finding the optimal reference state that maximizes the accessible entanglement
of a system in the presence of the local particle number SSR. Under such a SSR, the pure state describing a single particle
which shared coherently between two sites is physically equivalent to an equal mixture of states representing a single
particle at one site and a single particle at the other. In general, the SSR constrains the accessible entanglement due to the
unobservabilty of the coherences between eigenkets of different local particle number. The accessible entanglement in
this case is the weighted average of the entanglement found in each subspace of fixed local particle number. The
entanglement lost because of the SSR can be recovered with a reference ancilla and it is this problem that we focus on
here. In particular, we examine cases where the number of particles in the reference ancilla is not arbitrarily large and we
explore how effectively the reference ancilla increases the accessible entanglement in a variety of situations. In Section
\ref{C} we calculate the optimal reference state for maximizing the accessible entanglement from a system containing a
single shared particle where the number of particles in the reference system is fixed. We compare the amount of
 entanglement made accessible by the optimal state with that due to various other states in section \ref{E} and make steps
towards finding optimal reference state for a general state of the system in section \ref{D}.

\section{Optimal reference state for a single shared particle}\label{C}

\subsection{Particle entanglement}

We begin by briefly recalling the definition of the particle entanglement (or equivalently, the accessible entanglement) as
defined by Wiseman and Vaccaro \cite{HMW}.   As shown above, SSRs can be induced by the lack of a shared reference
or a conservation law. The origin of the SSR is unimportant for the main results of this paper.  However, as the
conservation of particle number was used in Ref.~\cite{HMW}, we assume the conservation also holds here. The
observation of coherences between states of different particle number requires operations that do not conserve particle
number. The conservation of particle number therefore implies that phase coherences between subspaces of the Hilbert
space corresponding to different numbers of particles are unobservable. This means that phase shifts generated by
$\exp(-i\hat N\theta)$, where $\hat N$ is the particle number operator and $\theta$ a phase angle, are not detectable. The
group of undetectable transformations is therefore the group U(1)$=\{\exp(-i\hat N\theta): 0\le \theta<2\pi\}$ of phase
shift operators, which induces the {\it particle number superselection rule} U-SSR. In the case of bipartite systems, we
consider the SSR which is induced by imposing the conservation of particle number at each spatial site.  The
corresponding group is $\U\AB(1)=\U\A(1)\otimes \U\B(1)$ where $\U_\mu(1)$ is the group of operators that
generate phase shifts at site $\mu$.   We call this the {\it local} particle number superselection rule, i.e. the {\it local}
U-SSR.

Consider the pure state $\ket{\Psi ^{(N)}}_{\rm AB}$ of a system comprising a fixed number of indistinguishable
particles shared between two spatially separated sites labeled $A$ and $B$. We have included a superscript in the label
of the state to indicate the total (fixed) number of particles the state represents.  The effective state under the local
particle number SSR is not $\ket{\Psi ^{(N)}}_{\rm AB}$ but rather a mixed state given by \cite{BartlettWiseman}
\beqa
                   \hat\rho^{(N)} &=& \int_{2\pi} \frac{d\theta_A}{2\pi}\int_{2\pi} \frac{d\theta_B}{2\pi}
                      \left[e^{-i(\hat N_A\theta_A+\hat N_B\theta_B)}\ket{\Psi ^{(N)}}_{\rm AB}\right]\nn\\
                      &&\times\left[{}_{AB}\bra{\Psi ^{(N)}}e^{i(\hat N_A\theta_A+\hat N_B\theta_B)}\right]\\
                   &=& \sum_{n=0}^N \hat{\Pi}^{(n,N-n)}\AB\ket{\Psi ^{(N)}}_{\rm AB}
                 \bra{\Psi ^{(N)}}\hat{\Pi}^{(n,N-n)}\AB\
\eeqa
where $\hat{\Pi}^{(n,N-n)}\AB$ is a operator which projects onto the subspace representing $n$ particles at site A and
$(N\!-\!n)$ at B. The entanglement of $\hat\rho^{(N)}$ is, by definition, the particle entanglement of $\ket{\Psi
^{(N)}}_{\rm AB}$. We note that the effective state, $\hat\rho^{(N)}$, is a mixture of a set of mutually orthogonal,
entangled pure states $\hat{\Pi}^{(n,N-n)}\AB\ket{\Psi ^{(N)}}_{\rm AB}$. Its entanglement is thus the average of the
entanglement of each member of the set. This means that the particle entanglement is given by \cite{HMW}
\beq
    \label{E_P}
                E_{\rm P}(\ket{\Psi^{(N)}}_{\rm AB}) \equiv
          {\sum}_n p_n E_M(\ket{\Psi^{(n)}_{\rm proj}}_{\rm AB}) \ ,
\eeq
where
\beqa
    \ket{\Psi^{(N)}_n}_{\rm AB}
                    &=& \frac{\hat{\Pi}^{(n,N-n)}\AB\ket{ \Psi\N } _{ \rm AB}}{ \sqrt{p_n} }\ ,
                    \label{proj}\\
          p_n &=& {} _{ AB} \bra{\Psi\N } \hat{\Pi}^{(n,N-n)}\AB \ket{ \Psi\N }_{ AB} \ ,\nn\\
          E_M(\ket{\Psi^{(N)}_n}_{\rm AB}) &=& S({\rm Tr}_A\left[\ket{\Psi^{(N)}_n}_{\rm AB}\bra{\Psi^{(N)}_n}\right])
         \  .\nn
\eeqa
Here  $S(\hat\varrho)$ is the von Neumann entropy of $\hat\varrho$, $\ket{\Psi^{(N)}_n}_{\rm AB}$ is the state of the
system after detecting $n$ particles at site A, and $E_M(\ket{\psi})$ is the entropy of entanglement of the state
$\ket{\psi}$.  In the current context  $E_M(\ket{\psi})$ corresponds to the entanglement of the spatial modes that are
occupied by the particles.

\subsection{Particle entanglement of system and reference ancilla}
Two of us previously showed \cite{HMW} that two copies of a shared single-particle system contained particle
entanglement whereas (as noted above) each of the systems does not. Evidently one system behaves as a reference for the
other. We further developed developed this concept in Ref.~\cite{VAWJ} where we showed that a shared particle ancilla
can be used as a reference to increase the accessible the entanglement in another system. We now examine the problem of
finding the state of the reference ancilla that yields the {\it maximum value} of the particle entanglement of the
combined system and reference ancilla for a given state of the system. We confine our analysis to situations where the
system and reference ancilla are in pure states to avoid unnecessary detail.  Also in accord with our assumption that the
particle number is conserved, we constrain the total number of particles in each of the system and the reference ancilla to
be fixed \footnote{Our constraint of a fixed number of particles means that it is not possible to make a direct
comparison between the reference states considered in this paper and van Enk's refbit state \cite{vanEnk} which is based
on an uncertain number of shared particles.}.  Our focus will initially be on the simplest case, that of a single shared
particle in the state:
\beq
            \ket{\Psi^{(1)}}\AB = \frac{1}{\sqrt{2}}(\ket{0,1}\AB+\ket{1,0}\AB)
            \label{singleparticle}
\eeq
where the ket $\ket{n,m}_{\rm AB}$ represents the occupation of spatial modes at the sites A and B by $n$ and $m$
identical particles, respectively, in second quantization notation.   It is straightforward to show that this state possesses
no particle entanglement, i.e. $E_P( \ket{\Psi^{(1)}}\AB)=0$ \cite{HMW,VacAnsWis}. Let the reference ancilla
consist of $M$ particles of the same type as the system and be prepared in the pure state

\beq
                \ket{\Phi\M}\AB=\sum _{n=0} ^{M} c_n \ket{n,M-n}\AB\ ,
         \label{Phi_M_ancilla}
\eeq
for which the coefficients $c_n$ satisfy the normalization condition
\beq
      \ip{\Phi\M}{\Phi\M}= \sum_{n=0}^M|c_n|^2 = 1
      \label{constraint}
\eeq
but are otherwise undetermined.  This ancilla state has been intentionally constructed to have no particle entanglement
itself, i.e. $E_P( \ket{\Phi\M}\AB)=0$. Any particle entanglement in the combined system and ancilla state, i.e. in
\beqa
       \ket{C^{(M+1)}}\AB &=& \ket{\Psi^{(1)}}_{\rm AB} \otimes \ket{\Phi\M}_{\rm AB}\ ,
\eeqa
will therefore be due to the ancilla's ability to ameliorate the effects of the local U-SSR.

Expanding $\ket{C^{(M+1)}}_{\rm AB}$ in terms of states with fixed local particle number yields
\beq
       \ket{C^{(M+1)}}  = \sum_{n=0}^{M+1} \sqrt{p_n} \ket{C_{n}^{(M+1)}}
\eeq
where $p_n=\ip{ C^{(M+1)} |\hat{ \Pi }^{(n,M+1-n)}_{AB}}{
C^{(M+1)} }$ and
\beq
       \ket{C_{n}^{(M+1)}} = \frac{1}{\sqrt{p_n}}\hat{\Pi}^{(M+1,n)}_{AB}\ket{C^{(M+1)}}  \label{continue}
\eeq
is a state of representing a total of $n$ particles at site A and $(M+1-n)$ at site B.  For convenience,  we omit the site
labels AB here and below when they are implied by the context.  The projection operator
$\hat{\Pi}^{(n,M+1-n)}_{AB}$ is of the same basic form as the one in \eqr{proj} but here it projects onto the subspace
of states of the combined system plus ancilla with a total of exactly $n$ particles at A and $(M+1-n)$ at B. That is, it
distinguishes between the sites but it is insensitive to the individual modes at each site. Making use of
\eqr{singleparticle} we find that
\beqa
     \ket{C_{n}^{(M+1)}} &=& \frac{1}{\sqrt{2p_n}}\Big( c_{n-1} \ket{1,0}\otimes\ket{n-1,M-n+1} \nn\\
     &&\ \ \ \ \ + c_n\ket{0,1}\otimes\ket{n,M-n}\Big)
\eeqa
for $n=0,1,\ldots,M+1$.  To make expressions more compact here and in the following, we have introduced two extra
coefficients $c_{-1}$ and $c_{M+1}$ whose values are zero, i.e.
\beq
         |c_{-1}|^2=|c_{M+1}|^2=0 \label{boundary_conditions} \ .
\eeq
We shall refer to \eqr{boundary_conditions} as the ``boundary conditions''. As before, the value of $p_n$ ensures that
$\ip{C_{n}^{(M+1)}}{C_{n}^{(M+1)}}=1$ and is easily found to be given by
\beq
           p_n =     \frac{1}{2}(|c_{n-1}|^2 + |c_{n}|^2)  \label{p_n_1sh_particle}
\eeq
for $0\le n\le M+1$. Taking the partial trace of $\ket{C_{n}^{(M+1)}}\bra{C_{n}^{(M+1)}}$ over a basis
representing states at site B yields
\beqa
        && \frac{|c_{n-1}|^2}{2p_n}\Big(\ket{n-1}\otimes\ket{1}\bra{1}\otimes\bra{n-1}\Big) \nonumber \\
                  &+&\frac{|c_{n}|^2}{2p_n}\Big(\ket{n}\otimes\ket{0}\bra{0}\otimes\bra{n}\Big)
           \ . \label{Tr_B_ancilla_system_1_sh_particle}
\eeqa
The entanglement of modes $E_M(\ket{C_{n}^{(M+1)}})$ is given by the von Neumann entropy of
\eqr{Tr_B_ancilla_system_1_sh_particle}, i.e.
\beqa
          E\m(\ket{C_{n}^{(M+1)}}) &=& -\frac{1}{2p_n}|c_{n-1}|^2\log_2(\frac{|c_{n-1}|^2}{2p_n}) \nonumber \\
                   &&\ \ \ \ - \frac{1}{2p_n}|c_n|^2\log_2(\frac{|c_n|^2}{2p_n}) \ ,
\eeqa
and so, from \eqr{E_P}, the particle entanglement is the
average
\beqa
    &&\hspace{-9mm}E\p( \ket{C^{(M+1)}} ) = \sum _{n=0} ^{M+1} p_n E\m(\ket{C_{n}^{(M+1)}})\nn\\
           &=& \frac{1}{2\log_2 e}\sum _{n=0} ^{M+1} \Big[-2|c_{n}|^2\ln(|c_{n}|^2) \nn\\
           &&\ \ \   + (|c_{n-1}|^2+|c_{n}|^2)\ln(|c_{n-1}|^2 +|c_n|^2)\Big]\ .\nn\\
          \label{E_P_1sh_particle}
\eeqa

\subsection{Conditions for the optimum reference}

The maximization of the particle entanglement in \eqr{E_P_1sh_particle} over the coefficients $c_n$ is  subject to the
normalization of the ancilla state $\ket{\Phi^{(M)}}$ given in \eqr{constraint}. We note that the optimization can be
performed with respect to $|c_n|^2$ rather than $c_n$ since the particle entanglement \eqr{E_P_1sh_particle} and the
constraint \eqr{constraint} are both functions of $|c_n|^2$ only. Let
\beq
                    F=(2\log_2 e) E\p(\ket{C^{(M+1)}})-\alpha(\sum _{n=0} ^M |c_n|^2-1)
               \label{g}
\eeq
be the auxiliary function where $ \alpha $ is a Lagrange multiplier.

The coefficients of the optimal reference state are given by the extremum of \eqr{g} and so satisfy
\beqa
                 \frac{\partial F}{\partial |c_n|^2}
                &=& -2[\ln(|c_n|^2)+1]+[\ln(|c_{n-1}|^2+|c_n|^2)+1]\nn\\
                &&\ \ \ +[\ln(|c_n|^2+|c_{n+1}|^2) +1]-\alpha\nn\\
          &=& \ln\Big[\frac{(|c_{n-1}|^2+|c_n|^2)(|c_n|^2+|c_{n+1}|^2)}{|c_n|^4}\Big]-\alpha\nn\\
          &=&0 \label{h}
\eeqa
for $n=0,1, \ldots, M$. This equation can be expressed in terms of a {\em recurrence relation} as follows
\beq
        \beta |c_n| ^4 = |c_n| ^4 + |c_{n-1}| ^2 |c_{n+1}| ^2 + |c_ n |^2 (|c_{n-1}| ^2 + |c_{n+1}| ^2)
               \label{recurrence}
\eeq
 for $n=0,1,\ldots,M$ where we have set $\beta
= e^\alpha$.  The boundary conditions in \eqr{boundary_conditions} ensure the recurrence relation has the correct form
for $n=0$ and $n=M$.

To find the solution it is convenient to first rearrange the recurrence relation as
\beq
          |c_{n+1}|^2 = \frac{(\beta-1)|c_n|^4 - |c_ n |^2|c_{n-1}|^2}{|c_ n |^2+|c_{n-1}|^2}\ .
    \label{recurrence_up}
\eeq
Iterations of this starting from $n=0$ lead to explicit expressions for the coefficients in which $|c_n|^2$ is a polynomial
in $\beta$ of order $n$. The details are given in the Appendix. Unfortunately, the derivation of a solution from these
expressions that satisfies the upper boundary condition $|c_{M+1}|^2=0$ and the normalization condition
\eqr{constraint} does not appear tractable analytically. Nevertheless we note that the system state \eqr{singleparticle} is
symmetric with respect to the interchange of labels, $\ket{n,m}\AB\mapsto \ket{m,n}\AB$, at each site A and B, which
implies that if there is a {\em unique} optimal reference state, it will also be symmetric under the same operation. Indeed
we also show in the Appendix using the polynomial expressions that
\beq
               |c_n|^2=|c_{M-n}|^2
         \label{symmetry_condition}
\eeq
for $n=0,1,\ldots,M$. This result will be useful later.

\begin{figure}
\begin{center}
\epsfig{file=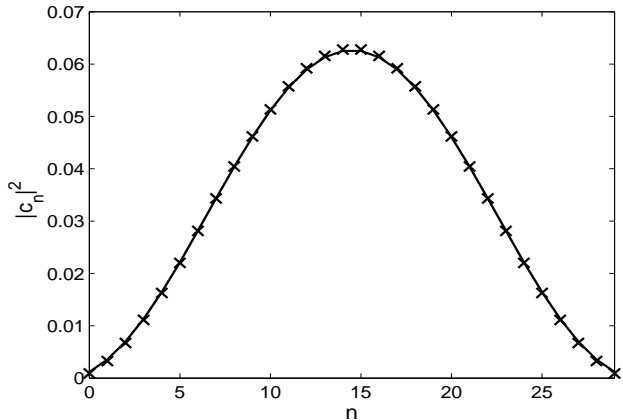, height=6cm,width=9cm} \caption{Optimal
reference state for a system comprising a single shared particle.
The probability $|c_n|^2$ is plotted as a function of the number of
particles $n$ at site A for the numerical solution (crosses) and our
ansatz (solid curve) for an ancilla with a total of $M=29$.
 For the ansatz we used the values
$A=1$ and $\epsilon=\frac{3}{2}$ which are correct to $O(M{}^{-2})$.
 }\label{num_soln_1_sh_particle}
\end{center}
\end{figure}

The optimal reference state for any given value of $M$ can be determined numerically by computing the values of
$\beta$ and the set of coefficients $\{|c_n|^2\}$ that satisfy the recurrence relation \eqr{recurrence_up}, the
normalization condition \eqr{constraint} and the boundary conditions \eqr{boundary_conditions}. The approach we
adopted was as follows: (i) make an initial guess of the value of $\beta$,  (ii) use the lower boundary condition
$|c_{-1}|^2=0$ and the recurrence relation \eqr{recurrence} to compute the values of the variables
$|c'_n|^2=|c_n|^2/|c_0|^2$  for $n=0,1,\ldots,M$, (iii) use the computed values of $|c'_n|^2$ and the normalization
condition \eqr{constraint} to determine the values of the normalized coefficients $|c_n|^2$. This yields a set of
normalized coefficients $|c_0|^2,|c_1|^2\ldots,|c_M|^2$ that depend on the value of $\beta$. (iv) adjust the value of
$\beta$ using a bisection method to find the zero of $|c_{M+1}|^2$ and repeat steps (ii)-(iv) until a desired tolerance is
reached.   In Fig.~\ref{num_soln_1_sh_particle} we plot the numerically determined solution to the recurrence relation
\eqr{recurrence} for $M=29$. The figure clearly illustrates the predicted symmetry about $n=(M+1)/2$.

\begin{figure}
\begin{center}
\epsfig{file=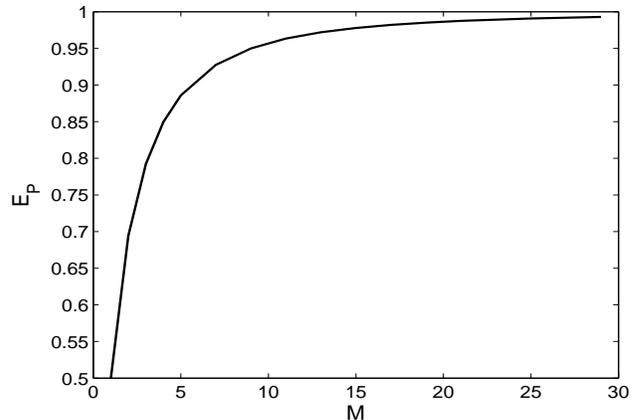, height=6cm,width=9cm} \caption{The
amount of particle entanglement $E_P$ versus the number of particles
in the reference ancilla, $M$, for a system consisting of a single
shared particle. The ancilla is in the optimal reference state
discussed in the text.}\label{epopt}
\end{center}
\end{figure}

In Fig.~\ref{epopt} we plot the particle entanglement, $E_P(\ket{C^{(M+1)}})$, of the combined system plus ancilla
against the total number of particles, $M$, in the ancilla. The fact that $E_P(\ket{C^{(M+1)}})>0$ whereas
$E_P(\ket{\Psi^{(1)}})=E_P(\ket{\Phi^{(M)}})=0$ shows that the ancilla partially shields the entanglement of the
system from the local U-SSR.

\subsection{An ansatz for the optimal reference}

The numerical solution depicted in Fig.~\ref{num_soln_1_sh_particle} suggests that the analytical solution may be of a
trigonometric form. To check, we now consider the ansatz for the solution to the recurrence relation of the form
\beq
              |c_n|^2 = B^{-1} \{ A-\cos[z(n)] \}
        \label{general}
\eeq
where
\beq
           z(n)=\xi n + \theta \ .
     \label{z_n}
\eeq
Here $A$, $B$, $\xi$ and $\theta$ are constant for a given value of $M$, i.e. a given number of particles in the ancilla.
We note that the numerical solution in Fig.~\erf{num_soln_1_sh_particle} comprises almost one full cycle of the
cosine function. This implies $\xi$ can be expressed conveniently in the form
\beq
            \xi=\frac{2\pi}{M+2\epsilon}
      \label{xi_epsilon}
\eeq
where $\epsilon$ is an adjustable parameter. Moreover the symmetry condition \eqr{symmetry} implies a condition on
the value of $\theta$ as follows.  Replacing the left and right sides of \eqr{symmetry_condition} using the ansatz
\eqr{general} and the expression for $\xi$ in \eqr{xi_epsilon} shows
\beq
               \cos \Big[\frac{2\pi}{M+2\epsilon}n+\theta \Big]=\cos \Big[\frac{2\pi}{M+2\epsilon}(M-n)+\theta
         \Big] \ ,
\eeq
which is satisfied by
\beq
              \theta=\frac{2\pi\epsilon}{M+2\epsilon}\ .
        \label{theta}
\eeq
Hence the final form of our ansatz, given by Eqs.~\erf{general}-\erf{theta} is
\beq
               |c_n| ^2 = B^{-1} \bigg\{A - \cos \Big[ \frac{ 2 \pi (n+ \epsilon )}{M+2\epsilon } \Big] \bigg\}
         \label{ansatz}
\eeq
and depends on just three parameters $A$, $B$ and $\epsilon$.

We now check the validity of the ansatz by substituting it into the recurrence relation \eqr{recurrence}. To simplify
matters we use the form given by Eqs.~\erf{general} and \erf{z_n}. The result of the substitution is
\beqa
     &&\hspace{-10mm} (\beta - 1)\Big\{A - \cos [z(n)] \Big\}^2  \nonumber \\
                 &=&  \Big\{ A-\cos[z(n+1) ] \Big\} \Big\{ A-\cos[z(n-1) ] \Big\} \nonumber \\
                 &&\ \ \ +\Big\{A- \cos [ z(n) ] \Big\}\nn\\
           &&\ \ \ \times\Big\{2A- \cos [z(n+1)]-\cos[z(n-1)] \Big\} \ , \nn\\
           \label{recurrence_with_ansatz}
\eeqa
which is independent of the normalization constant $B$.  Using trigonometric identities and grouping terms it is easy to
show that the right hand side of this equation can be expressed as
\beqa
     {\rm RHS} &=&  3A^2 +\cos^2(\xi)-1+\big[ 1+2 \cos(\xi) \big]    \nonumber \\
         &&\times\big\{-2A \cos[z(n)]+\cos^2[z(n)]\big\}\ .
                      \label{n}
\eeqa
Similarly the left hand side of \eqr{general} is
\beq
              {\rm LHS} = (\beta -1)\Big\{ A^2 - 2A \cos[z(n)]+\cos^2[z(n)]\Big\} \ .
        \label{o}
\eeq
Comparing the coefficients of $\cos[z(n)]$ and $\cos^2[z(n)]$ in Eqs.~\erf{n} and \erf{o} shows that the value of
$\beta$ that satisfies the recurrence relation is given by
\beq
         \beta - 1= 2 \cos(\xi) + 1 \label{beta=cos(xi)}\ .
\eeq
Comparing the constant terms in these equations similarly shows that $ (\beta - 1 )A^2 = 3A^2 -\sin^2(\xi) $. These last
two results imply that
\beq
        A^2 = \frac{ \sin^2(\xi)}{2[1-\cos(\xi)]}=\frac{1+\cos(\xi)}{2} \ . \label{p}
\eeq
Eliminating $\cos(\xi)$ and $\sin(\xi)$ in \eqr{p} using \eqr{beta=cos(xi)} then yields
\beq
      \beta=4A^2\ .  \label{beta=4A^2}
\eeq
These results show that the ansatz does indeed satisfy the recurrence relation for particular values of the parameters $A$,
$B$ and $\epsilon$.

To determine the actual values of the parameters we need to apply additional conditions. We first consider the boundary
conditions $|c_{-1}|^2=0$ and $|c_{M+1}|^2=0$. Setting $n=-1$ in the detailed form of the ansatz \eqr{ansatz} yields
\beq
               B^{-1} \left\{A - \cos \left[ \frac{ 2 \pi (\epsilon-1)}{M+2 \epsilon }\right] \right\}=0 \ ,
\eeq
which gives an expression for $A$ in terms of $\epsilon$ as
\beq
                 A = \cos \left[ \frac{ 2 \pi (\epsilon-1)}{M+2 \epsilon } \right] \ .
           \label{A_epsilon_1}
\eeq
The same result is obtained for $n=M+1$ due to the symmetry condition \eqr{symmetry_condition}. Finally we note
that the analytic solution $|c_1|^2=|c_0|^2(\beta-1)$ given in \eqr{general} yields another expression involving $A$ and
$\epsilon$ as follows. Using the ansatz \eqr{ansatz} to replace the coefficients $|c_0|^2$ and $|c_1|^2$ and
\eqr{beta=4A^2} to replace $\beta$ gives
\beq
       A-\cos\left[ \frac{ 2 \pi (1+\epsilon)}{M+2 \epsilon } \right] =\left[A-\cos(\frac{ 2 \pi \epsilon}{M+2 \epsilon})
                   \right](4A^2-1) \ .
                \label{A_epsilon_2}
\eeq
This analysis shows that our ansatz, which depends on the values of just three parameters $A$, $B$ and $\epsilon$,
satisfies all the conditions for the optimal reference state provided Eqs.~\erf{A_epsilon_1} and \erf{A_epsilon_2} are
satisfied. Indeed the values of the parameters $\epsilon$ and $A$ are determined by solving Eqs.~\erf{A_epsilon_1} and
\erf{A_epsilon_2} simultaneously, and the value of $B$ determined from the normalization condition \eqr{constraint}.
Once these values are determined, our ansatz \eqr{ansatz} provides an {\em exact analytical expression} for the optimal
reference state for any value of $M$.

We can also find approximate values for $A$, $B$ and $\epsilon$ in the large $M$ regime. Substituting for $A$ using
\eqr{A_epsilon_1} and simplifying yields
\beqan
    &&\sin\left(\frac{2\pi\epsilon}{M+2\epsilon}\right)\sin\left(\frac{2\pi}{M+2\epsilon}\right)\\
                &&\quad \quad =\sin\left[\frac{2\pi(\epsilon-\frac{1}{2})}{M+2\epsilon}\right]\sin\left(\frac{2\pi\frac{1}{2}}{M+2\epsilon}\right)\\
         &&\hspace{3cm}\times \left\{4\cos^2\left[ \frac{ 2 \pi (\epsilon-1)}{M+2 \epsilon } \right]-1\right\}
\eeqan
which, in the large $M$ regime, gives
\beq
      \epsilon=\frac{3}{2} + O(M^{-2}) \ .
\eeq
Similarly, we find from \eqr{A_epsilon_1} in the same regime that
\beq
       A= 1 +  O(M^{-2})\ .
\eeq
The normalization constant $B$ is found by evaluating $\sum_n|c_n|^2=1$ using the ansatz \eqr{ansatz}, i.e.
\beq
      B=\sum_{n=0}^{M} \left\{A - \cos \left[ \frac{ 2 \pi (n+ \epsilon )}{M+2\epsilon } \right] \right\}\ .
\eeq
For the case where $M$ is odd we find, using the symmetry condition \eqr{symmetry_condition}, that
\beq
         B=2\sum_{n=0}^{\frac{M-1}{2}}f(n)
\eeq
where
\beq
         f(x) =  \left\{A - \cos \left[ \frac{ 2 \pi (x+ \epsilon )}{M+2\epsilon } \right] \right\}
\eeq
is a monotonically increasing function over the range from $x=0$ to $x=\frac{M+1}{2}$.  We note that
\beq
         2\int_{0}^{\frac{M+1}{2}}dx f(x) \ge B \ge 2\int_{0}^{\frac{M+1}{2}}dx f(x-1)\ ,
\eeq
that is,
\beq
         A(M+1)+1+\epsilon \ge B \ge A(M+1) -3+\epsilon\ ,
\eeq
and so
\beq
        B^{-1}= \frac{1}{A(M+1)} + O(M^{-2})\ .
\eeq
The same result is also found for the case where $M$ is even. Hence in the large $M$ regime the analytical form of the
optimal reference state is
\beqa
                 |c_n|^2 &=& \frac{1}{M+1}\left\{1 - \cos \left[ \frac{ 2 \pi (n+ 3/2 )}{M+3} \right] \right\} + O(M^{-2})\ \nn\\
                   &=& \frac{2}{M+1}\sin^2\left[\frac{\pi (n+ 3/2 )}{M+3} \right] + O(M^{-2})\ .
              \label{ansatz coefficients}
\eeqa

For comparison, in Fig.~\ref{num_soln_1_sh_particle} we have also plotted  as the solid curve the values of the
probabilities $|c_n|^2$ given by the ansatz in \eqr{ansatz} for $M=29$ for the values $\epsilon=\frac{3}{2}$ and
$A=1$, which are correct to $O(M^{-2})$. The figure shows that the ansatz with these approximate values of
$\epsilon$ and $A$ is in relatively close agreement with the exact numerical solution (crosses) at this value of $M$.

\subsection{General nature of the optimal reference states}\label{subsec:general nature}

To see why the optimal reference state has the general form shown in
Fig.~\ref{num_soln_1_sh_particle} and \eqr{ansatz coefficients}
consider two reference states in opposing limiting cases. First, in
the limit of a narrow particle number distribution only one
coefficient is non-zero; it is straightforward to show using
\eqr{E_P_1sh_particle} that the particle entanglement
$E_P(\ket{C^{(M+1)}})$ is zero for this reference state. In the
other extreme, the particle number distribution is uniform with
$|c_n|^2=1/(M+1)$ for $n=0,1,\ldots,M$; the right hand side of
\eqr{E_P_1sh_particle} is $1-M/(M+1)$ for this reference state.
Clearly a reference system with broad particle number distribution
has an advantage for making particle entanglement accessible.
However, the reason for $E_P(\ket{C^{(M+1)}})$ being less than 1 bit
in the broad particle number distribution case can be traced to the
boundary conditions \eqr{boundary_conditions}.  This suggests that a
distribution which vanishes in near the boundaries $n=0$ and $n=M$
also has an advantage. Evidently the particle number distribution of
the optimal reference state balances these two opposing effects,
breadth of the distribution with vanishing values near the
boundaries, to maximize $E_P(\ket{C^{(M+1)}})$.

Our derivation of the optimal state fixes only the modulus of the
coefficients $c_n$ in Eq. (2.6). The arguments of the complex
numbers, $c_n$, are completely arbitrary. This is somewhat
surprising given the role of the ancilla is presumed to act as a
phase reference. We now explore how the ability to act as a
reference for the U-SSR differs from the ability to act as a phase
reference. First note that the asymmetry of a state depends solely
on the modulus of number state coefficients $c_n$ \cite{VAWJ}. For
example, consider the state of the reference ancilla representing a
uniform sharing of particles with arbitrary complex arguments given
by \beq
      \ket{\Phi^{(M)}}\AB =\frac{1}{\sqrt{M+1}}\sum _{n=0}^M e^{i \theta _n} \ket{n,M-n}\AB
\eeq where $\{\theta_n :0\le \theta_n< 2\pi\}$ are an unordered set
of phase angles. Under the local group operation
$\hat{T}\A(k\Delta)=e^{-i\hat N\A k\Delta}\otimes \id\B$ where
$\Delta=2\pi/(N+1)$ this state transforms to \beq
      \frac{1}{\sqrt{M+1}}\sum _{n=0}^M e^{i(\theta _n-n\phi_k)}\ket{n,M-n}\AB
\eeq which is orthogonal to $\ket{\Phi^{(M)}}\AB$  for integer $k$
satisfying $1\le k\le N$. This shows that $\ket{\Phi^{(M)}}\AB$ is
{\it asymmetric} with respect to $\U(1)$ and thus breaks the local
$\U(1)$ symmetry \footnote{In contrast, the state $\ket n$ is
invariant to the group operation $\hat{T}(\phi)$ in the sense that
$\hat{T}(\phi)\ket n\bra n\hat{T}^\dagger(\phi)=\ket n\bra n$, which
illustrates the {\it symmetry} of $\ket n$ with respect to $\U(1)$
\cite{VAWJ}.}. As this result is independent of the values of the
phase angles $\{\theta_n\}$ it confirms that only the {\it moduli of
the number state coefficients} are important in terms of the
U(1)-SSR. Moreover, \eqr{E_P_1sh_particle} shows that the particle
entanglement of a system consisting of one shared particle and an
ancilla in reference state $\ket{\Phi^{(M)}}\AB$ is independent of
the phases, $\{\theta_n\}$, of the reference state.

\begin{figure}
\begin{center}
\epsfig{file=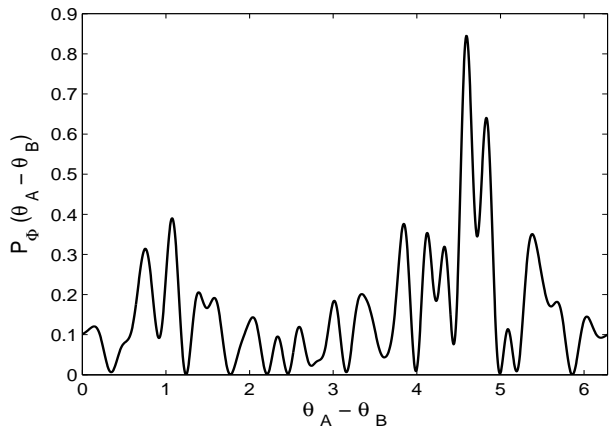, height=6cm,width=9cm} \caption{The
distribution of the relative phase difference for a random
distribution of $\{\theta_n \} $.}\label{rand}
\end{center}
\end{figure}

\begin{figure}
\begin{center}
\epsfig{file=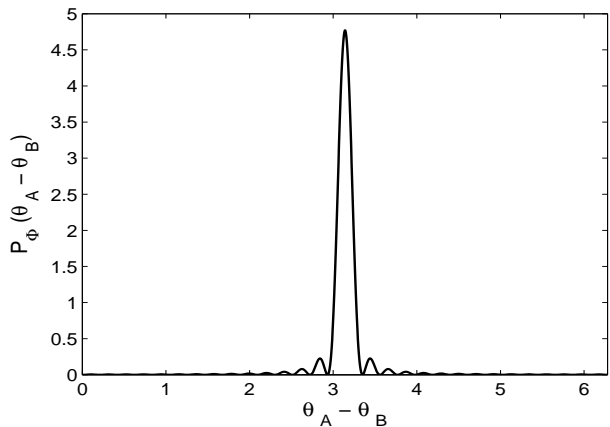, height=6cm,width=9cm} \caption{The
distribution of the relative phase difference where $\{\theta _n\}$
is linear in $n$.}\label{nonrand}
\end{center}
\end{figure}

Next consider the phase properties of the state
$\ket{\Phi^{(M)}}\AB$ for various choices of the set of phase angles
$\{\theta_n\}$. For this we use the Pegg-Barnett phase formalism for
physical states in the infinite-$s$ limit
\cite{phase1,phase2,phase3}. The joint phase probability density
$P_\Phi(\theta\A,\theta\B)$ for phase angles $\theta\A$ and
$\theta\B$ which describe the phase operators $\hat\phi\A$ and
$\hat\phi\B$ of the ancilla spatial modes at sites A and B,
respectively, is given by  \beq
    P(\theta\A,\theta\B) = \frac{1}{(2\pi)^2}\left|\sum_{n,m} e^{i(n\theta\A+m\theta\B)}\ip{n,m}{\Phi^{(M)}}\AB\right|^2
    \label{Phase_prob_density_joint}
\eeq For the reference state $\ket{\Phi^{(M)}}\AB$ this can be
rewritten as \beqa
    P(\theta\A,\theta\B) =  \frac{1}{2\pi}{\cal P}_\Phi(\theta\A-\theta\B)
\eeqa where \beqa
                          {\cal P}_\Phi(\theta\A-\theta\B)&=& \frac{1}{2\pi(M+1)}
                          \left|\sum_{n=0}^M e^{i[n(\theta\A-\theta\B)+\theta_n]}\right|^2\nn\\
                          \label{Phase_difference_prob_density_general}
\eeqa is the probability density for the phase difference
$\hat\phi\A-\hat\phi\B$ and the factor  $1/2\pi$ in
\eqr{Phase_prob_density_joint} is the uniform phase probability
density for the phase operator of either spatial mode. The highly
correlated nature of the phase difference stems from the sharing of
a fixed number $M$ of particles between the two modes. Figure
\ref{rand} shows this phase distribution for the case where $\theta
_n = \pi n$ whereas figure \ref{nonrand} shows the phase
distribution where $\theta _n$ is random. Note that both phase
distributions equally alleviate the U-SSR. Clearly there is a {\it
fundamental difference between the ability to act as a phase
reference and the ability to act as a reference for the U-SSR}, viz.
the complex phase of the number state coefficients are crucial for
the former but are unimportant for the latter. While this conclusion
applies to the specific case of a single-shared particle system as
considered in \eqr{E_P_1sh_particle} , we show later in Section
\ref{D} that it also extends to arbitrary systems states.

\section{Comparison of the Optimal Reference Frame to Other States}\label{E}

States with broad particle number distributions are known to have well-defined phase properties
\cite{SummyPegg,phase1,phase2,phase3}. So it is interesting to compare our optimal reference states with classes of
states that have been optimized for other phase dependent quantities.  We define a figure of merit, $D$, based on the
excess particle entanglement that is made accessible by the optimal reference state as follows:

\beq
       D = \frac{ E_{\rm p}[ \ket{C^{(M+1)}_ {\rm opt}}  ] -  E_{\rm p}[ \ket{C^{(M+1)}  } ] }{ E_{\rm p}[ \ket{ C^{(M+1)}_{\rm opt }} ] }\ ,
               \label{definition of delta}
\eeq
where
\beqa
           \ket{C^{(M+1)}_ {\rm opt}}&=& \ket{\Psi^{(1)}}\otimes\ket{\Phi^{(M)}_ {\rm opt}}\\
       \ket{C^{(M+1)}}&=&\ket{\Psi^{(1)}}\otimes\ket{\psi^{(M)}}\ ,
\eeqa
for single particle system state $\ket{\Psi^{(1)}}$, optimal
reference ancilla state $\ket{\Phi^{(M)}_ {\rm opt}}$ and arbitrary reference ancilla state $\ket{\psi^{(M)}}$. We now
consider a number of states whose phase properties are important in some way. All the states are constructed on the
Hilbert space ${\mathbb H}_{\rm R}$ which is spanned by the number state basis $\{\ket{n,M-n}: n=0,1,\ldots,M\}$.

The {\it Berry-Wiseman} phase optimized states of a two-mode optical field give the optimum phase shift estimation in
a Mach-Zehnder apparatus for a fixed total number of particles (photons) under ideal canonical phase measurements
\cite{BerryWiseman}. These states have the form
\beq
                  \ket{\psi } \propto \sum _{n=0}^M \sin[
            \frac{\pi(n + \epsilon )}{M+2 \epsilon}] \ket{n,M-n} \label{phase}
\eeq
where $M$ is the number of particles and  $\epsilon=1$.  The {\it Summy-Pegg}  phase optimized states of a single
mode optical field give the minimum phase variance for a fixed upper bound in the particle (photon) number distribution
\cite{SummyPegg}. Their two-mode version on the Hilbert space ${\mathbb H}_{\rm R}$ is given by \eqr{phase} with
a parameter value of $\epsilon \approx 0.84$ for $M\gtrsim 10$. The variance of the optimized phase quantity for both
classes of states scales as $1/M^2$. A class of states with less phase resolution is given by the single-mode {\it coherent
states} $\ket\alpha\propto \sum_n \alpha^n\ket n/\sqrt{n!}$ where $\ket n$ is the usual number state.  These states have
a phase variance which scales as $1/|\alpha|^2$ and are optimized in the sense that they approximate number-phase
minimum uncertainty states \cite{phase2} for $|\alpha|\gg 1$. To make a comparison with our optimal reference state we
construct a two-mode version in the Hilbert space ${\mathbb H}_{\rm R}$ with an analogous number state expansion
to $\ket\alpha$ as follows
\beq
                  \ket{ \phi } \propto \sum _{n=0}^M \sqrt{\frac{(\frac{M}{2})^n}{n!}} \ket{n,M-n} \ .
\eeq
The parameter that is analogous to the field amplitude $\alpha$ in $\ket\alpha$ has the value $\sqrt{M/2}$ here. Thus in
the large $M$ regime the variance of the two-mode phase operator scales as $2/M$.  We also consider reference states
whose particle number probability distribution $|c_n|^2$ corresponds to a binomial distribution, i.e.
\beq
            \ket{\psi }= \sum _{n=0}^M \left[\left(
           \begin{array}{c} M \\ n
                  \end{array} \right) p^n(1-p)^{(M-n)}\right]^{1/2}\ket{n,M-n}
\eeq
for which we set $p=0.5$ to make the distribution symmetric. These states are a two-mode generalization of Stoler {\em
et al.}'s single mode {\it binomial states} \cite{stoler}. Finally we include in our comparison the {\it shared phase state},
\beq
                 \ket{\psi }= \frac{1}{\sqrt{M+1}}\sum _{n=0}^M \ket{n,M-n}\ ,
\eeq
which is a two mode version of the single-mode Pegg-Barnett phase state with zero phase \cite{phase1, phase3}. This
state belongs to the class of states on ${\mathbb H}_{\rm R}$ with maximum asymmetry with respect to the particle
number SSR \cite{VAWJ}.  The change in the von Neumann entropy of these states due to the action of the SSR is the
maximum value of $\log_2 M$ and so these states play an important role in terms of breaking the symmetry represented
by the SSR \cite{VAWJ}.

Fig.~3 shows that the figure of merit, $D$, of all states, at least
after a certain value of $M$, are monotonically decreasing with $M$.
However, the decrease in $D$ for increasing $M$ for the
Berry-Wiseman and the two-mode Summy-Pegg minimized phase variance
states is far greater than the two-mode versions of the coherent and
binomial states. The reason for this can be traced to our earlier
observation in Sec. \ref{subsec:general nature} that broad
particle-number distributions are an advantage in making the
entanglement accessible.  In particular the former two states have a
particle-number standard deviation that scales as $M$ whereas the
two-mode coherent states have a particle-number standard deviation
that scales as $\sqrt{M}$. Compared to the maximum possible width
$M$ of the distribution, the former are relatively broad and the
latter is relatively narrow. We note that the figure of merit $D$
for the coherent and binomial states become closer as $M$ increases
as expected from the fact that the associated particle-number
distributions approach each other as $M\to\infty$.

\begin{figure}
\begin{center}
\epsfig{file=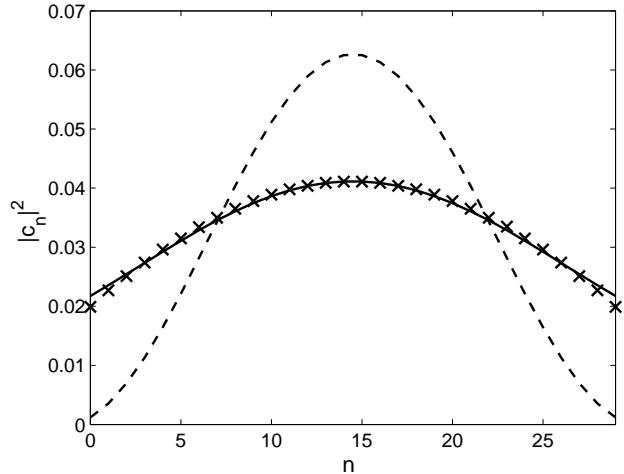, width=9cm}\caption{A comparison of the figure
of merit $D$ for the shared phase states ($\circ$), two-mode
versions of the coherent states (*), binomial states ($+$) and
Summy-Pegg phase optimized states ($\times$), and Berry-Wiseman
phase optimized states ($\Box$). $D$ represents the relative
effectiveness of a given reference state compared to the optimal
reference state. The effectiveness of all states increases as the
total number of particles $M$ increases for large $M$.
}\label{fig:comparison}
\end{center}
\end{figure}

\section{Optimal Reference states for general pure states}\label{D}

We now generalize our results to the case where there are $N$ particles in the system and, as before, $M$ particles in the
ancilla. Consider the tensor product of two arbitrary states of the system and ancilla for this case given by
\[
         \ket{C^{(M+N)}}\AB = \ket{\Psi\N}\AB \otimes \ket{\Phi\M}\AB\ ,
\]
where the system and ancilla states are
\beq
         \ket{\Psi\N}\AB = \sum _{n=0}^N d_n
            \ket{n, N-n}\AB\ ,
\eeq

\beq
       \ket{\Phi\M}\AB = \sum _{n=0}^M c_n \ket{n, M-n}\AB \ ,
              \label{Phi_M_ancilla_N_M}
\eeq
respectively.  The coefficients $c_n$ and $d_n$ are subject to the normalization conditions
\beq
        \sum _{n=0}^N |d_n|^2 = \sum _{m=0}^M |c_m|^2 = 1 \ .
               \label{norms_c_n_d_n}
\eeq
Neither the system nor the ancilla contain entanglement of
particles i.e. $E\p(\ket{\Psi\M}\AB)=E\p(\ket{\Phi\M}\AB)=0$. To simplify the notation we shall omit the subscripts A
and B from here onwards. The state of the combined system can be expressed as
\beqa
        \ket{C^{M+N}} &=& \sum _{n=0}^N\sum _{m=0}^M  d_n c_m \ket{n,N-n}\otimes \ket{m,M-m} \nn\\
      &=&\sum_{k=0}^{N+M}\sqrt{p_k}\ket{C_k^{(M+N)}}
\eeqa
where $\ket{C_k ^{(M+N)}}$ is a state containing exactly $k$
particles at site A and $(N+M-k)$ at site B. Also $p_n$ is a
normalization constant, i.e.
\beqa
                \ket{C_{k}^{(M+N)}}
                   &=&\frac{1}{\sqrt{p_k}}\sum _{n=0}^N\sum _{m=0}^M \delta_{n+m,k}d_n c_m\nn\\
                   &&\ \ \ \times\ket{n,N-n}\otimes\ket{m,M-m} \\
          p_k&=&\sum _{n=0}^N\sum_{m=0}^M|d_n c_m|^2 \delta _{n+m,k}\ .\label{p_k_N_M_case}
\eeqa
The calculation of $E\p$ in Section \ref{C} for a system comprising one particle is easily extended to the case $N$
particles as follows. We find that
\beqa
               &&E\p(\ket{C^{(M+N)}}) = \sum_{k=0}^{N+M} p_k E\m(\ket{C_k^{(M+N)}})\nn\\
              &&= - \sum_{k=0}^{N+M} p_k \sum_{n=0}^N\sum_{m=0}^M\frac{|d_n c_m|^2 }{p_{k}}\delta _{n+m,k}
              \log_2 \frac{|d_n c_m|^2 }{p_{k}}\nn \\
        &&= - \sum_{k=0}^{N+M}\sum_{n=0}^N\sum_{m=0}^M|d_n c_m|^2 \delta _{n+m,k}
        \log_2 \frac{|d_n c_m|^2 }{p_{k}}  \ .\label{E_P_N_M}
\eeqa We again find that the particle entanglement is independent of
the complex phases of the number state coefficients $c_n$ of the
reference state.  This shows that our conclusion in Section
\ref{subsec:general nature} for the special case of a single
shared-particle system state, namely that a phase reference and a
reference for the U-SSR are fundamentally different, applies also to
the general case of arbitrary system states.

\subsection{The infinite-particle ancilla state}\label{infinite}

One may suspect that in the limit of $M\to\infty$, the ancilla could resemble a classical reference and completely shield
the system from the local particle number SSR.  We now show this using the uniformly distributed ancilla state with
$c_n=1/\sqrt{M+1}$ for $M\gg N$. The state with exactly $k$ particles at site A for $N\le k\le M$ is given by
\beqa
         &&\ket{C^{(M+N)}_{k}}\nn\\
       &&\quad=\frac{1}{\sqrt{p_k}}\sum _{n=0}^N d_n\ket{n,N-n}\otimes\ket{(k-n),M-(k-n)}\nn
\eeqa
and so
\beq
         E\p(\ket{C^{(M+N)}_{k}})=E\m(\ket{C^{(M+N)}_{k}})\ ,
\eeq
and the probability $p_k$ in \eqr{p_k_N_M_case} of finding this
state is simply
\beq
        p_k=\sum _{n=0}^N\frac{|d_n |^2}{M+1}=\frac{1}{M+1}\ .
\eeq
The contribution to $E\p(\ket{C^{(M+N)}})$ from these states is
\[
        X=\sum _{k=N}^M p_{k} E\m (\ket{C^{(M+N)}_{k}}) = \frac{ M+1-N}{M+1} E\m(\ket{\Psi\N}) \ .
\]
It is straightforward to show that the remaining $2N$ states $\ket{C_k^{(M+N)}}$ for $0\le k<N$ and $M< k\le
N+M$ have $E\m (\ket{C^{(M+N)}_{k}})\le E\m(\ket{\Psi^{(N)}})$ and $p_k\le1/(M+1)$ and so their contribution
to $E\p(\ket{C^{(M+N)}})$ is less than
\[
        Y=\frac{2N}{M+1} E\m(\ket{\Psi\N})\ .
\]
These last two results give bounds on $E\p(\ket{C^{(M+N)}})$ as
\[
         X \le  E\p(\ket{C}) \le X+Y\ .
\]
In the limit $M\to\infty$, $X\to E\m(\ket{\Psi\N})$ and $Y\to 0$ and so
\[
        E\p(\ket{C^{(M+N)}})\rightarrow E\m(\ket{\Psi\N})\ .
\]
The infinite-particle ancilla effectively shields the system from the particle number SSR and makes all of its
entanglement accessible. We note that the class of perfect reference ancillae for the SSR include states that are poor
phase references.

\subsection{Conditions for general optimal ancilla states}

We now use the Lagrange multiplier method to find conditions for the coefficients $c_n$ of the optimal ancilla state for
an arbitrary, but fixed, value of $M$.  We want to maximize the particle entanglement $E\p(\ket{C^{(M+N)}})$ over
$|c_n|^2$ subject to the normalization constraint $\sum_{n=0}^M |c_n|^2 = 1$. The optimal ancilla state for any system
state is given by the extremum of the auxiliary function
\beq
        F=E\p(\ket{C^{(M+N)}}) - \alpha_1( \sum_{n=0}^M |c_n|^2 - 1)
\eeq
where $E\p(\ket{C^{(M+N)}})$ is given by \eqr{E_P_N_M} and
$\alpha_1$ is the Lagrange multiplier. Setting the derivative
$\partial F/\partial |c_m|^2$ to zero we find with a little effort
that
\beq
         \alpha _1 = \sum _{n=0} ^N \Big[-|d_n|^2 \log_2 (|d_n c_m|^2)+ |d_n|^2 \log_2 p_{n+m}  \Big]
                \label{t}
\eeq
for $m=0,1,\ldots M$ and where $p_k$ is given by \eqr{p_k_N_M_case}.

It is easy to show that the symmetry condition \eqr{symmetry_condition} does not hold in general. Consider, for
example, the simplest case given by $N=M=1$, where both the system and ancilla contain a single shared particle.
Substituting these values into Eq. \erf{t} yields for $m=0$
\[
                  \alpha _1 = |c_1|^2 \log_2 \frac{|c_1d_0|^2+|c_0d_1|^2}{|c_0d_0|^2}
\]
and for $m=1$
\[
                  \alpha _1 = |c_0|^2 \log_2 \frac{|c_1d_0|^2+|c_0d_1|^2}{|c_0d_1|^2} \ .
\]
The simultaneous solution of these equations is given by
\[
                  \frac{ 2^x - 1}{x} = \bigg| \frac{d_0}{d_1} \bigg| ^2 \ ,
\]
where $x=| c_1/c_0| ^2 $.  This implies that the optimum reference state for a  non-symmetric system state,
$|c_0|\ne|c_1|$, is also non-symmetric with $|d_0|\ne |d_1|$.

The analytical treatment of system and ancilla states with arbitrary values of $N$ and $M$, respectively, is beyond the
scope the current paper. Instead we focus here on the situation where the system state is of a simple form, namely, a
shared phase state with $d_n=1/\sqrt{N+1}$, i.e. we set
\beq
         \ket{\Psi\N} = \frac{1}{\sqrt{N+1}} \sum_{n=0}^{N} \ket{n,N-n}\ \label{conditionone} .
\eeq
We further restrain the problem to the case where $N=M$ to make the problem tractable.

A shared phase state has a symmetry between the sites in the sense that it is invariant under an interchange or site labels,
i.e. $AB \mapsto BA$. The corresponding optimal reference state has the same symmetry with coefficients of the form
\beq
         |c_m|^2 = |c_{M-m}|^2
                \label{c_n_is_c_M-m_for_N_M}
\eeq
for $m=0,1,\ldots,M$. The proof of this is as follows. We find
from Eqs.~(\ref{t}) and (\ref{p_k_N_M_case}) the set of equations
\beq
         \alpha_1 = -\log_2 (\frac{1}{N+1}|c_m|^2)+ \frac{1}{N+1}\sum _{n=0} ^N \log_2 p_{n+m}
                \label{alpha_1}
\eeq
for $m=0,1,\ldots,M$ where
\beq
              p_k=\frac{1}{N+1}\sum_{j=k-N}^k|c_j|^2 \ .
\eeq
Here, for compactness, we have extended the set of coefficients
$c_n$ with the values $c_n=0$ for $n<0$ and $n>M$. Taking the
exponential of both sides of \eqr{alpha_1} yields

\beqa
         \beta|c_m|^{2(N+1)} = \prod_{n=0}^N \Big(\sum_{j=n+m-N}^{n+m}|c_j|^2\Big)
                \label{c_m_for_N_M}
\eeqa
Omitting the zero terms $|c_j|^2$ for $j<0$ and $j>M$ and
setting $N=M$ we find {\arraycolsep 0pt
\beqa
                 \beta|c_m|^{2(M+1)} &=& \prod_{n=0}^{M-m} \Big(\sum_{j=0}^{n+m}|c_j|^2\Big)
         \!\prod_{n=M-m}^{M}\! \Big(\sum_{j=n+m-M}^{M}\!|c_j|^2\Big)\nn\\
           &=&\prod_{n=m}^{M} \Big(\sum_{j=0}^{n}|c_j|^2\Big)
      \prod_{n=0}^{m}
      \Big(\sum_{j=n}^{M}|c_j|^2\Big) \ ,
        \label{trimmed c_m_for_N_M}
\eeqa
where we have made use of the normalization
\eqr{norms_c_n_d_n} to include an extra factor in the first line.
 Using this result we find the relationship between the coefficients $|c_m|^2$ for sequential values of the index $m$ is
 }
\beqa
                 \beta|c_{m+1}|^{2(M+1)} &=& \beta|c_m|^{2(M+1)}\frac{1-\sum_{j=0}^{m}|c_j|^2}{\sum_{j=0}^{m}|c_j|^2}\nn\\
                      \label{sequential_up}\\
             \beta|c_{m-1}|^{2(M+1)}&=& \beta|c_m|^{2(M+1)}\frac{1-\sum_{j=m}^{M}|c_j|^2}{\sum_{j=m}^{M}|c_j|^2}\ .\nn\\
                 \label{sequential_dn}
\eeqa
We also find by setting specific values of $m$ in \eqr{trimmed
c_m_for_N_M} that
\beqa
         \beta|c_0|^{2(M+1)} &=& \prod_{n=0}^{M} \Big(\sum_{j=0}^{n}|c_j|^2\Big)\\
                 \beta|c_M|^{2(M+1)} &=& \Big(\sum_{j=0}^{M}|c_j|^2\Big)\prod_{n=0}^{M-1} \Big(\sum_{j=n}^{M}|c_j|^2\Big)
\eeqa
and so $|c_0|^2=|c_M|^2$.  Using this result together with Eqs.~(\ref{sequential_up}) and (\ref{sequential_dn}) with
the specific values of $m=0$ and $m=M$, respectively, shows that $|c_1|^2=|c_{M-1}|^2$. Repeating this analysis for
sequential values of $m$ completes the proof of the symmetry in \eqr{c_n_is_c_M-m_for_N_M}.

Another property of the optimal state is that the distribution $|c_m|^2$ is unimodal which can be seen as follows. First
we define the difference $\Delta_{m+1,m}$ as
\beq
                  \Delta _{m+1,m}=|c_{m+1}|^{2(M+1)}-|c_{m}|^{2(M+1)}
\eeq
which, from \eqr{sequential_up}, is given by
\beqa
                 \Delta _{m+1,m} &=& |c_{m}|^{2(M+1)}(\frac{1-2\sum _{j=0}^{m} |c_j|^2}{\sum _{j=0}^{m} |c_j|^2}) \label{dcn} \ .
\eeqa
From the symmetry property \eqr{c_n_is_c_M-m_for_N_M} of the
optimal state and the normalization condition \eqr{norms_c_n_d_n} we know that
\beq
                  \sum _{m=0}^{[M/2]} |c_m|^2=\sum _{m=[M/2]+1}^M |c_n|^2=\frac{1}{2} \label{evensum}
\eeq
where $[M/2]$ is the largest integer satisfying $[M/2]\le M/2$. Next we use this result to determine the positivity of the
differences $\Delta_{m+1,m}$. Consider if the expression $1-2\sum _{j=0}^{m} |c_j|^2$ in \eqr{dcn} were zero for a
value of $m$ less than $[M/2]$. This would imply that $|c_{m+1}|^2,|c_{m+2}|^2,\ldots,|c_{[M/2]}|^2=0$ and by
\eqr{dcn} that $\Delta_{m+1,m}=0$, which would mean that $|c_{m}|^2=|c_{m+1}|^2=0$ and so $1-2\sum
_{j=0}^{m-1} |c_j|^2=0$. It follows, by induction, that the coefficients $|c_j|^2$ would be zero for
$j=0,1,\ldots,[M/2]$ which contradicts \eqr{evensum}. Hence $1-2\sum _{j=0}^{m} |c_j|^2=0$ only for $m=[M/2]$
and so from \eqr{dcn} $\Delta_{j+1,j}>0$ for $j=0,1,\ldots,([M/2]-1)$. Taking account of the symmetry
\eqr{c_n_is_c_M-m_for_N_M} then shows that the distribution $|c_n|^2$ is monotonically increasing over the index
$m$ for $m=0,1,\ldots,[M/2]$ and monotonically decreasing for $m=[M/2],([M/2]+1),\ldots,M$. This completes the
proof of the unimodal property.

\begin{figure}
\begin{center}
\epsfig{file=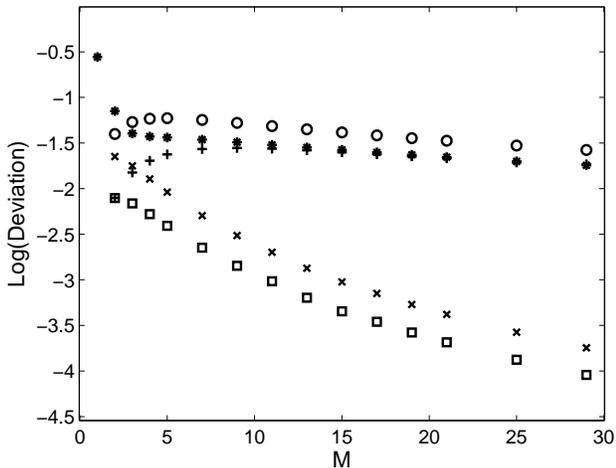, width=9cm}\caption{Comparison of the particle number
distributions for the optimal reference state $\ket{\Phi^{(M)}_{\rm opt}}$
($\times$) and the trial state $\ket{\phi^{(M)}_{\rm trial}}$  (---) for $N=M=29$.
The comparison shows that the optimal reference state is very well approximated by the trial state. Moreover, a
comparison with the optimal state for $N=1,M=29$ ($-\;-$) shows
that the optimal particle number distribution broadens out as the number of particles in the system, $N$,
increases. Finally we note that distribution for the optimal reference state ($\times$) satisfies the symmetry and unimodal properties derived
analytically. }\label{comparison}
\end{center}
\end{figure}

In summary, the optimal state for a system in the shared phase state \eqr{conditionone} and a reference ancilla with an
equal maximum number of particles, i.e. with $M=N$, has coefficients $|c_m|$ that are symmetric about a modal point
and are monotonically increasing before this point and are monotonically decreasing after it. These properties are
reminiscent of the ansatz used for a single shared particle case which suggests that the optimal state would be well
approximated by a {\it trial state}  $\ket{\phi^{(M)}_{\rm trial}}$ which is of the form \eqr{ansatz} for a suitable
choice of the parameters $A$ and $\epsilon$, and a range of $M$ values. Exact analytical solutions or better
approximations are beyond the scope of this paper. However, one can solve the $N=M$ case numerically using either
\eqr{sequential_up} or \eqr{sequential_dn}. Fig.~\ref{comparison} shows the particle number distribution for the exact
optimal reference state with $N=M=29$ compared with that of the  trial state for the parameter values $A=1.9$ and
$\epsilon = 8.9$.  The figure confirms our intuition that the exact solution in the $N=M$ case can be approximated by a
 state which is of the form \eqr{ansatz}. The inner product
$\ip{\Phi^{(M)}_{\rm opt}}{\phi^{(M)}_{\rm trial}}$ of the exact solution with the trial state differs from one by
$~6\times10^{-6}$.

\section{Discussion}

We have focused on the situation where the ancilla state $\ket{\Phi\M}$ has a particularly simple form, as given by
\eqr{Phi_M_ancilla} and \eqr{Phi_M_ancilla_N_M}, corresponding to a single spatial mode at each site. Such states
automatically satisfy the requirement $E\p(\ket{\Phi\M})=0$, i.e. that the ancilla does not contain particle entanglement.
Nevertheless our analysis generalizes quite easily to situations with multiple spatial modes and multiple ancillae. The
most general pure ancilla state of $M$ particles which contains no particle entanglement has the form
\beq
       \ket{\Phi\M}=\sum_{m=0}^M c_m \ket{\phi (m)}_{\rm A}\otimes\ket{\varphi (M-m)}_{\rm B}
              \label{general_ancilla_state}
\eeq
where $\ket{\chi (n)}_\mu$, for $\chi = \phi \ {\rm or} \
\varphi$, is an arbitrary normalized state of exactly $n$ particles at site $\mu$ in the second quantization formalism.
For example, if the part of the ancilla at site $\mu$ involves the 3 mutually orthogonal spatial modes $\ket{\cdot}_\mu
\otimes \ket{\cdot}_\mu \otimes\ket{\cdot}_\mu$, then the general expression for $\ket{\chi(n)}_\mu$ is
\beq
               \ket{\chi(n)}_\mu=\sum_{i,j,k}f_{i,j,k}\ket{i}_{\rm A}
                 \otimes \ket{j}_{\rm A}\otimes\ket{k}_{\rm A}\delta_{i+j+k,n}
           \label{arbitrary_local_ancilla_state}
\eeq
where $f_{i,j,k}$ are normalized coefficients.  Notice that the
set of states at each site are orthonormal, viz. $_\mu\ip{\chi
(n)}{\chi (m)}_\mu=\delta_{n,m}$. To incorporate these more general
ancilla states in our previous analysis we need only replace the
ancilla states $\ket{n}_\mu$ in the above with
$\ket{\phi_\mu(n)}_\mu$.  However, as our analysis used only the
orthogonality property of the set of $\ket{n}_\mu$ and not its
detailed structure, the coefficients $|c_m|^2$ for the optimal
ancilla state would be unchanged.  Hence the solutions derived above
represent classes of optimal ancilla states for arbitrary choices of
coefficients of the kind $f_{i,j,k}$ in
\eqr{arbitrary_local_ancilla_state}.

Moreover consider an ancilla which comprises component subsystems of
the form
\beq
         \ket{{\Phi'}\M}=\ket{\Phi^{(M_1)}}\otimes\ket{\Phi^{(M_2)}}
                \label{tensor_product_form}
\eeq
where $M=M_1+M_2$.  The requirement that
$E\p(\ket{{\Phi'}\M})=0$ implies that $\ket{{\Phi'}\M} $ can be expressed in the form
\beq
       \ket{{\Phi'}\M}=\sum_{m=0}^M c'_m \ket{\phi (m)}_{\rm A}\otimes\ket{\varphi (M-m)}_{\rm B}
              \label{tensor_product_ancilla_state}
\eeq
where the tensor product form of \eqr{tensor_product_form} restricts the variability of the coefficients $c'_m$.  This is
a further constraint on the coefficients, in addition to the normalization condition, and so the entanglement made
accessible by a composite ancilla cannot exceed that given by a single component ancilla composed of a maximum of M
particles. Hence the use of multiple ancillae in a tensor product state does not give any advantage in optimizing the
accessible entanglement.

\section{Conclusion}

Quantum reference frames are of particular theoretical interest and the phenomena of unlocking entanglement has
proven to be a very interesting application. Our analytical techniques have proven to be fruitful in producing exact
solutions in the case of a single shared particle and guidelines for sensible approximations for system states of large
$N$. Much work is needed to produce exact analytical solutions for more general system states. However, the results
extracted seem to establish the intuition that pure states with a broad particle-number distribution act as good reference
frames for the particle number SSR. Indeed, the optimal reference state in the infinite-particle limit $M \rightarrow
\infty$ makes all the entanglement in the single shared particle accessible.

\acknowledgements{This work was supported by the Australian Research
Council and the Queensland State Government.}

\appendix
\section{Solution as a series of polynomials in $\beta$}\label{polyapp}
Here we explore the nature of the iterative solution of the recurrence relation \eqr{recurrence_up} for the case where the
system consists of a single particle which is equally shared between the sites. Setting $n=0$ and using the boundary
condition $|c_{-1}|^2 = 0$ gives an expression for $|c_1|^2$ in terms of $|c_0|^2$. Next, by setting $n=1$ we find an
expression for $|c_1|^2$ and so on. Continuing in this way yields a solution to Eq. \erf{recurrence_up} that is expressed
in terms of just $\beta$ and $|c_0|^2$ as
\beq
               |c_n|^2 = P_n(\beta) |c_0|^2 \label{poly_C_n_is_P_C_o}
\eeq
where $P_n(\beta)$ are polynomials in $ \beta $ of order $n$ given by
\beqa
         P_0(\beta) &=& 1 \nn \\
        P_1(\beta) &=&   \beta-1 \nn \\
        P_2(\beta) &=&   \left( \beta-1 \right)  \left( \beta-2 \right) \nn \\
        P_3(\beta) &=&   \left( {\beta}^{2}-3\,\beta+1 \right)  \left( \beta-2 \right)\nn \\
        P_4(\beta) &=&   \left( \beta-3 \right)  \left( \beta-1 \right)  \left( {\beta}^{2}-3\,\beta+1 \right)\nn \\
        P_5(\beta) &=&  \left( {\beta}^{3}-5\,{\beta}^{2}+6\,\beta-1 \right)  \left( \beta-3 \right)  \left( \beta-1 \right)\nn \\
        P_6(\beta) &=&  \left( {\beta}^{3}-6\,{\beta}^{2}+10\,\beta-4 \right)  \left( {\beta}^{3}-5\,{\beta}^{2}+6\,\beta-1 \right)\nn\\
\eeqa
{\em etc.} Due to the normalization condition \eqr{constraint}, we have the property that
\beq
              \sum_{n=0}^M P_n(\beta) = \frac{1}{|c_0|^2} \ .
\eeq
Alternatively, rearranging the recurrence relation \eqr{recurrence} as
\beq
               |c_{n-1}|^2 = \frac{(\beta-1)|c_n|^4 - |c_ n |^2|c_{n+1}|^2}{|c_ n
              |^2+|c_{n+1}|^2} \ ,
        \label{recurrence_dn}
\eeq
leads to a different form of the solution.  For this we set $n=M$ and use the boundary condition $c_{M+1}=0$ with
\eqr{recurrence_dn} to get an expression for $|c_{M-1}|^2$ in terms of $|c_M|^2$.  Continuing in this way with
reducing values of $n$ we find that any coefficient $|c_{M-n}|^2$ can be written in terms of polynomials in $\beta$ and
$|c_M|^2$. The symmetry of \eqr{recurrence} with respect to interchanging $n$ with $m=M-n$ ensures that the
polynomials are the same as those appearing in \eqr{poly_C_n_is_P_C_o}, i.e.
\beq
               |c_{M-n}|^2 = P_n(\beta) |c_M|^2 \ . \label{poly_C_M_minus_n_is_P_C_M}
\eeq
Substituting $n=M$ into \eqr{poly_C_M_minus_n_is_P_C_M} and \erf{poly_C_n_is_P_C_o}  yields
\beq
               |c_0|^2 = P_M(\beta ) |c_M|^2\label{P_M_eqn1}  \ ,     |c_M|^2 = P_M(\beta) |c_0|^2
\eeq
respectively. The simultaneous solution to Eqs.~\erf{P_M_eqn1} is
\beqa
              P_M(\beta)&=& 1\  , \
                      |c_0|^2 = |c_M|^2 \label{C_0_equals_C_M}
\eeqa
and so from Eqs.~\erf{poly_C_n_is_P_C_o} and \erf{poly_C_M_minus_n_is_P_C_M} we find that
\beq
                    |c_n|^2= |c_{M-n}|^2
         \label{symmetry}
\eeq
for $n=0,1,\ldots,M$. Thus the solution which represents the optimal reference state is symmetric in this sense.

\section{Phase properties of the state $\ket{\Phi^{(M)}_1}\AB$}\label{kerr}
The unitary operator  $\hat U\A=e^{i\hat N\A(\hat N\A-1) \vartheta}$
in the definition of the state $\ket{\Phi^{(M)}_1}\AB$ in
\eqr{ref_state_with_Kerr} represents a Kerr-like nonlinear
interaction.  The action of this operator on coherent states has
been discussed extensively in the literature (see e.g. Refs.~
\cite{MilburnKerr,BB_Kerr,MTK_Kerr,MT_Kerr,Vaccaro_Kerr}). However,
in our case the operator $\hat U\A$ acts on the shared phase state
$\ket{\Phi^{(M)}_01}\AB$ in \eqr{ref_state_before_Kerr} and so
requires a minor extension of previous work. For this we use a
previous result to derive a form of the operator $\hat U\A$ that
allows us to infer the pertinent phase properties of
$\ket{\Phi^{(M)}_1}\AB$.

One of us has previously shown that \cite{Vaccaro_Kerr} \beqa
    \sum_{k=0}^{K-1}c_k\exp(in\phi_k) = \exp[-i\vartheta n(n-1)]\\
    c_j = \frac{1}{K}\sum_{n=0}^{K-1}\exp[-i\vartheta n(n-1)]\exp(in\phi_j)
\eeqa
where
\beqa
    \vartheta &=& \frac{\pi J}{K}\\
    \phi_k &=& \phi_0 + \frac{2\pi k}{K}\\
    \phi_0 &=&\{[J(K-1)]\;\text{mod} 2\}\frac{\pi}{K}
\eeqa and $J$ and $K$ are positive integers sharing no common
factors.  Expanding $\hat U$ in terms of the Fock basis then yields
\beqa
  \hat U\A &=& \sum_{n=0}^\infty \exp[-i\vartheta n(n-1)]\ket{n}\A{}\A\!\bra{n}\\
             &=& \sum_{n=0}^\infty \sum_{k=0}^{K-1}c_k \exp(in\phi_k)\ket{n}\A{}\A\!\bra{n}\\
             &=& \sum_{k=0}^{K-1} c_k \exp(i\hat N\A\phi_k)
\eeqa which is in the form of a series of phase shifting operators
$\exp(i\hat N\phi_k)$.  The case discussed in the text is for the
choice $J=1$ and $K=2$ for which $\vartheta=\pi/2$, $\phi_0=\pi/2$,
$\phi_1=3\pi/2$, $c_0=(1+i)/2$, and $c_1=(1-i)/2$, and so
\setlength\arraycolsep{1pt} \beqa
      \hspace{-5mm}\ket{\Phi^{(M)}_1}\AB  &=&  \hat U\A\ket{\Phi^{(M)}_0}\AB \nn\\
      &=&\frac{1+i}{2\sqrt{M+1}}\sum _{n=0}^M e^{in\pi/2}\ket{n,M-n}\AB \nn\\
      && + \frac{1-i}{2\sqrt{M+1}}\sum _{n=0}^M e^{-in\pi/2}\ket{n,M-n}\AB\ .
\eeqa Using \eqr{Phase_difference_prob_density_general} then shows
that \eqr{Phase_prob_density for_kerr_state} is the phase difference
probability density corresponding to this state.

\bibliography{optstate}

\end{document}